\documentclass[12pt, a4paper]{article}
\usepackage{amsfonts, amssymb}
\usepackage{graphicx}
\usepackage{latexsym,amsmath,color,array}

\usepackage{enumerate}
\usepackage{bbding}
\usepackage{amssymb}
\usepackage{amsmath}
\usepackage{graphicx}
\usepackage{amsmath}
\usepackage{bbm}
\usepackage{mathtools}
 \usepackage{color}
 \usepackage{array}
 \usepackage{subfigure}
 \usepackage{multirow}
 \usepackage[dvipsnames]{xcolor}
 \usepackage{makecell}
 \usepackage{colortbl}

\def\1{\mathbb{I}}

\newcounter{thm}[section]
\newcounter{appen}[section]
\newcounter{assum}[section]

\begin{document}

\title{\Large \bf A Data-Driven Control-Theoretic Paradigm\\ for Pandemic Mitigation with Application to Covid-19}

\author{Kevin Burke\footnote{Corresponding author. University of Limerick, Ireland; kevin.burke@ul.ie}  \hspace{3cm}
B. Ross Barmish\footnote{Boston University and University of Wisconsin; bob.barmish@gmail.com.} }
\date{\today}

\maketitle

\begin{abstract}

In this paper, we introduce a new control-theoretic paradigm for mitigating the spread of a virus. To this end, our discrete-time controller, aims to reduce the number of new daily deaths, and consequently, the cumulative number of deaths. In contrast to much of the existing literature, we do not rely on a potentially complex virus transmission model whose equations must be customized to the ``particulars'' of the pandemic at hand. For new viruses such as Covid-19, the epidemiology driving the modelling process may not be well known and model estimation with limited data may be unreliable. With this motivation in mind, the new paradigm described here is data-driven and, to a large extent, we avoid modelling difficulties by concentrating on just two key quantities which are common to pandemics: the {\it doubling time}, denoted by~$d(k)$ and  the {\it peak day} denoted by~$\theta(k)$. Our numerical studies to date  suggest that our appealingly simple model can provide a reasonable fit to real data. Given that time is of the essence during the ongoing global health crisis, the intent of this paper is to introduce this new paradigm to control practitioners and describe a number of new research directions suggested by our current results.

\smallskip

\smallskip

\end{abstract}

\qquad

\newpage

\section{Introduction}\label{sec:intro}
\vspace{-.1in}

The main objective in this paper is to introduce a new control-theoretic paradigm for mitigation of the spread of pandemic disease. In the sequel, the discrete-time control variable $u(k)$ corresponds to the ``degree of mitigation'' which could be clinical or non-clinical in nature. The motivation for the paradigm to follow is simple to explain: When a new pandemic emerges on the world scene, standard virus transmission models are typically adopted in an effort to predict and control the uncertain future. However, with limited epidemiological information available for a novel virus, existing models may be inappropriate; i.e., such a virus may have many features in its transmission and mitigation-response dynamics which are not well understood and not captured using virus models from the past. Accordingly, with this motivation in mind, data-driven modelling is the focal point in this paper.

In the existing literature, the majority of papers on virus transmission are based on the so-called ``SIR'' compartmental model\cite{Kermack_McKendrick_1927} which, at a given point in time, places individuals into one of three disjoint classes: the \emph{susceptible} $S$ class, the \emph{infectious} $I$ class, and the \emph{recovered} $R$ class. The wide variety of extensions to this model include considerations such as death due to the disease, the latency time for infected individuals to become infectious, severity of illness, quarantining individuals, spatial and demographic effects, and social network structure; e.g., see the detailed accounts in \cite{Bailey_1975, Anderson_May_1991, Easley_Kleinberg_2010}. As stated in~\cite{Hethcote_1994}, there are ``a thousand and one epidemic models,'' and, therefore, customizing the model to the specifics of the situation at hand is non-trivial. 
This ``customization'' issue is epitomized by the variety of models even within the Covid-19 context, for example, in \cite{Giordano_2020,Gevertz_2020,Tuite_2020,Chen_2020}, we see a range of 7-16 possible states emphasizing different epidemiological considerations. In the early days of a pandemic, with limited data, not only is it difficult to build the structural model, but also to estimate its parameters. A particularly thorny modelling issue is that many infected individuals go undetected \cite{Li_2020}, in large part due to asymptomatic or mild cases. This runs counter to many SIR modelling efforts which assume that the \emph{detected} cases are equivalent to the \emph{true} cases when, in fact, the former will underestimate the latter.

%

Our approach in this paper, largely intended to counter concerns along the lines above, is motivated by data reliability issues which arise in the literature. To this end, our starting point is to work with data in the form of death numbers, which are typically higher fidelity as these are much less likely to be undetected or under-reported; e.g., see~\cite{Ma_2014}. Although at first sight, the SIR-related models of \cite{Mills_2004} and \cite{Flaxman_2020} also appear only to use death data, in fact, their models borrow infection and death-rate parameters from other papers in the literature. In another recent paper~\cite{Calafiore_2020}, this issue is partially remedied by using both death and case data with some additional parameters to account for the difference between detected and true cases.
In practice, the high variability in the detection process seems consistent with the wide uncertainty bands produced by \cite{Flaxman_2020} for the proportion of infected individuals in the population.

In summary, while research efforts aimed at understanding virus transmission mechanisms are valuable and important, the challenges in modelling will not be lost on the reader. Therefore, as mentioned above, we focus entirely on the death process data and propose a parsimonious three-parameter model. We expect that this approach will appeal to the control community wherein epidemiology is not a core competency. That being said, although we avoid transmission dynamics, we make use of two quantities which are common to epidemics
over limited time frames: the so-called ``doubling time'' and ``peak day'' for deaths; over longer horizons, we note that periodic peaking may occur~\cite{Hethcote_1991}. Thus, our approach can be viewed as lying between the epidemiologically-based SIR models and empirically-based phenomenological models in~\cite{Ma_2014} and~\cite{Chowell_2017}.

In addition to the considerations above, in this paper, we also include the dynamics of a
\emph{controller} reflecting the effect of mitigation measures aimed at reducing new deaths to zero. 
To the best of our knowledge, the earliest work bringing control-theoretic methodologies to epidemiological modelling dates back to the 1970s; see \cite{Gupta_1973,Hethcote_1973,Morton_1974}. Building on this, in~\cite{Behncke_2000}  the theoretical properties of various mitigation techniques are considered, while others focussed more specifically on vaccination \cite{Zaman_2008}, the combination of vaccination and treatment \cite{Yusuf_Benyah_2012}, non-clinical interventions such as social distancing, quarantining, and education \cite{Lin_2010}, system delay \cite{Zaman_2009,Briat_Verriest_2009} and spatiotemporal effects \cite{Rachik_2018};
more recently, the Covid-19 outbreak has been emphasized in \cite{Rachik_2020, Shorten_2020,Demasse_2020}. Interestingly, all of these control strategies assume underlying SIR-type models which, as previously discussed, present various challenges in practice. In contrast to these approaches, a key feature of our data-driven approach is the error dynamics involving the comparison of model-predicted deaths to actual deaths. This enables both the model and the mitigation controller to be adapted over time.

The remainder of the paper is organised as follows: In Section~\ref{sec:prelim} we describe the preliminaries involving the use of data and error dynamics at a high level. Subsequently, in Sections~\ref{sec:model} and~\ref{sec:openloop}, the details of our new model and its qualitative properties are provided for the case of constant mitigation. In Section~\ref{sec:numerical}, we describe the process of model parameter estimation and illustrate its use via numerical examples involving Covid-19 data from Brazil and Mexico, two of the countries among those with the highest of death rates as of mid-2020. Finally, in Section~\ref{sec:conclusions}, conclusions and directions for future research are described with emphasis being on issues of a control-theoretic nature.

\section{Preliminaries\label{sec:prelim}}
\vspace{-.05in}
In the sequel, we take~$k$ to be the index indicating the day number and~$u(k)$ to be the corresponding level of mitigation provided by the controller. In this first paper aimed at introducing our new paradigm, we do not consider the detailed mechanics of mitigation. Suffice it to say, large values of~$u(k)$ might represent stronger mitigation measures such as government mandates on social distancing and the use of masks and therapeutics whereas smaller values might correspond to ``relaxation'' of the rules.

Starting at $k=0$, at a general level, one begins with an equation for new deaths
$$
N(k+1) = f(D(k),N(k),u(k))
$$
where the cumulative total deaths are naturally constrained to be~$D(k) = D(k-1) + N(k)$. We propose an attractive form for~$f(D,N,u)$ in Section~\ref{sec:model}, based on the doubling time and peak day quantities from epidemiology.

\subsection{Data Driven Adaptive Mitigation}
\vspace{-.05in}
Our control-theoretic paradigm begins with the acquisition of daily death data which is collected to obtain estimates of the parameters  describing the 
function~$f$ above. Once the model is fixed, consistent with the tenets of receding horizon control, one can make predictions to determine if the control sequence~$u(k)$ is {\it mitigating} in the sense that~$\lim_{k \to \infty}N(k) = 0$.
In practice, the speed of convergence  is also a concern and strictly reaching the zero limit may not be required if the number of deaths is an acceptably small fraction of the population size.

\begin{figure}[htbp]
	\centering
	\includegraphics[width = 3.5in]{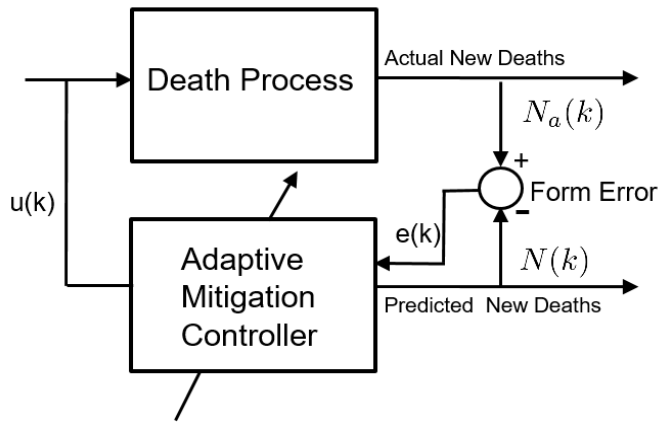}
	\caption{Adaptive Mitigation Control}
	\label{fig:simulation_01}	
\vspace{-.05in}
\end{figure}

As a pandemic unfolds over time, the model estimation should be updated periodically as new data is obtained. That is, letting~$N_a(k)$ denote the ``actual'' number of new deaths on day~$k$, as depicted in Figure~\ref{fig:simulation_01}, we use the  error~$e(k) \doteq N_a(k) - N(k)$ to drive the update of the model, the predicted deaths, and the associated adaptation of the controller~$u(k)$. There are many ways one could proceed when using the error; e.g., one can down-weight the distant past or smooth the noise using cumulative errors. However, such considerations are beyond the scope of this article.


\section{Model Equation Details\label{sec:model}}
\vspace{-.05in}
Our analysis begins with the so-called \emph{doubling time}, a widely reported metric used in practice to characterise the number of days~$d$ taken for the new deaths to double.\footnote{Note our use of the doubling time for new deaths whereas some authors and data providers use this terminology for cumulative total deaths.} If, for example, the doubling time is constant, 
the number of new deaths on day $k+1$ is given by $N(k+1) = 2^{1/d}N(k)$ so that~$N(k+d) = 2N(k)$ as expected. In practice, $d = d(k)$ is time-varying for reasons such as immunity building up in the population, changes to mitigation strategies, and medical developments. Therefore, with initial values $d(0) = d_0$ and $N(0) = N_0$, viewed as ``baseline'' quantities at the point from which we model, we obtain new deaths as
$$
N(k+1) = 2^{1/d(k)}N(k).
$$
In the equation above~$d(k)$ can be quite general in its functional form. We now specialise along the lines described in Section \ref{sec:intro}. That is, we structure~$d(k)$ so that it is consistent with many standard epidemic models which exhibit ``peaking'' behaviour. Thus, with $d_0 > 0$, the count on new deaths~$N(k)$ climbs until some {\it peak day}
$$
\theta(k) \doteq \theta(u(k))
$$
which may be time-varying due to changes in the level of mitigation characterized by the controller $u(k)$. After the peak day, $d(k)$ becomes negative so that we see declining~$N(k)$; in this post-peak regime, $|d(k)|$ may be viewed as the {\it halving time}. 

We capture the peaking behaviour by specifying the doubling model as
$$
d(k) = \frac{\theta(k)}{\theta(k) - k}d_0.
$$
which increases as $k$ approaches $\theta(k) > 0$ from the left. If, for some $k^*$, we have $k < k^* < \theta(k)$ and $k > k^* > \theta(k)$, then a single peak occurs at $k = k^* = \theta(k^*)$. More generally however, $\theta(k)$ may be referred to as the ``anticipated'' peak day since it can move further or closer in time due to the variations in $u(k)$ which could even yield multiple peaks.


\section{Solution for Open Loop Constant Control\label{sec:openloop}}
\vspace{-.05in}
An important starting point in this framework is the case when the level of mitigation is being held fixed over some extended time period. To this end, we consider the controller $u(k) = u_0$ for all~$k$. In practice, this would correspond to way the model is typically used; i.e, the mitigation level is held fixed between model updates and associated assessments of the efficacy of control measures in place. Subsequently, if the updated model parameters predict a worsening prognosis, decision makers might increase the level of mitigation~$u(k)$. For the ``unit step'' input control above, the corresponding peak day is denoted
$$
\theta_0 \doteq \theta(u_0).
$$
and we obtain the equation for new deaths
$$
N(k+1)  =  2^{1/d(k)}N(k) = 2^\frac{\theta_0 - k}{\theta_0 d_0}N(k)
$$
which is readily solved for~$k \ge 0$. Indeed, expressing~$N(k)$ as a product followed by summation of exponents yields
$$
N(k)  = \prod_{i=0}^{k-1}2^\frac{\theta_0 - k}{\theta_0d_0}N(0)
  = 2^{\frac{(2\theta_0+1)k - k^2}{2\theta_0d_0}}N_0.
$$

\subsection{Insights From New Death Solution}
\begin{figure}[htbp]
	\centering
	\includegraphics[width = 3.5in]{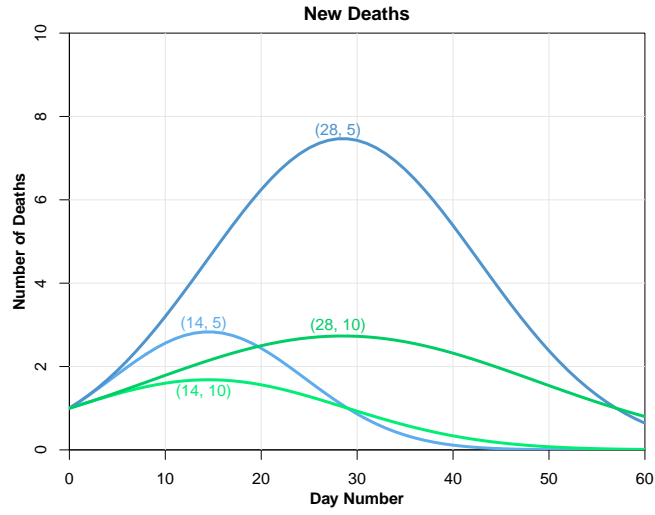}
	\caption{New Deaths Displayed for Four $(\theta_0,d_0)$ Cases with $N_0=1$}
	\label{fig:Nplot}
\vspace{-.05in}
\end{figure}

\vspace{-.05in}
The appealingly simple solution for~$N(k)$ above lends itself to various insights about the evolution and qualitative behavior of deaths. First, it is evident by inspection that~$N(k)$ is bell-shaped due to the negative $k^2$ exponent, and this is clear from Figure \ref{fig:Nplot}.
It is also straightforward to study the dependence of $N(k)$ on $\theta_0$ and $d_0$ by first noting that
\begin{align*}
\frac{\partial \log N(k)}{\partial \theta_0} &= \frac{k (k - 1)\log(2)}{2 \theta_0^2 d_0},\\
\frac{\partial \log N(k)}{\partial d_0} &= \frac{k(k-k_{N_0})\log(2)}{2 \theta_0 d_0^2}
\end{align*}
where~$k_{N_0} \doteq 2\theta_0+1$. For the most important case where $d_0$ and~$\theta_0$ are positive, we see that $N(k)$ increases in $\theta_0$, decreases in $d_0$ for $k < k_{N_0}$, and increases in $d_0$ for $k > k_{N_0}$.

It is apparent that $k_{N_0}$ represents the number of days until the pandemic reverts back to the early stage baseline level; i.e., $N(k_{N_0}) = N_0$. However, in contrast to the early stage, for~$k > k_{N_0}$, new deaths $N(k)$ are now on a downward trajectory which in some sense can be viewed as signalling the end point of the pandemic. Finally, it is also of interest to characterize the number of deaths on the peak day, 
$$
N(\theta_0) = 2^{\frac{\theta_0 + 1}{2 d_0}} N_0
$$
which, clearly, increases in $\theta_0$ and decreases in $d_0$.

\begin{figure}[htbp]
	\centering
	\includegraphics[width = 3.5in]{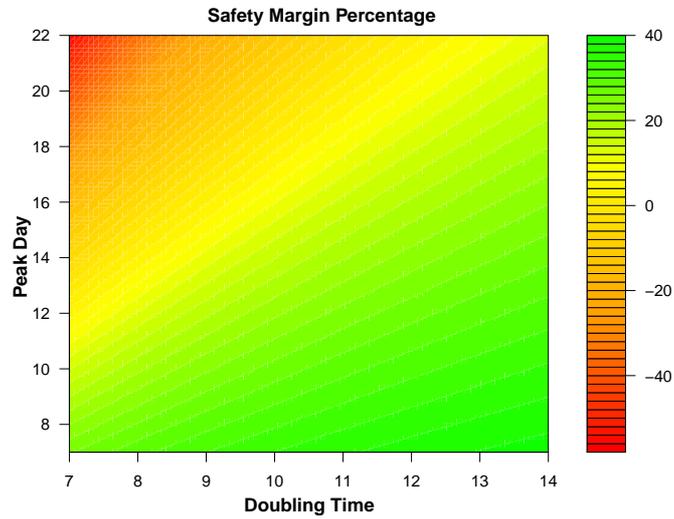}
\caption{Projected Safety Margin with $N_0/N_{max} = 0.5$}
	\label{fig:ICU}	
\vspace{-.05in}
\end{figure}

The peak~$N(\theta_0)$ above is important in that it correlates strongly with ``anticipated pressure'' on hospitals; e.g., it is a predictor of stress on resources such as intensive care units. In particular, given the concern that $N(\theta_0)$ exceeds some critical level~$N_{max}$, one can easily use the equations above to study {\it safety margins} such as
$$
S(d_0,\theta_0) = 100\left(1 - 2^{\frac{\theta_0 + 1}{2d_0}}
\frac{N_0}{N_{max}}\right)
$$
which indicates how far from~$N_{max}$, as a percentage, peak deaths will be. When $S < 0$, since this corresponds to $N(\theta_0) > N_{max}$, one might consider increasing the level of mitigation by updating~$u(k)$ to reduce a potential crisis. 
In Figure \ref{fig:ICU}, we display a realm of possible outcomes for this safety margin where, for example, with an initial doubling time of 10 days, an anticipated peak of up to about 14 days, the~$(d_0,\theta_0)$ point lies in the green-colored safety zone.

\subsection{Total Number of Deaths and Insights}
\vspace{-.05in}
The total number of deaths, given by 
$$
D(k) = \sum_{i=0}^k N(i),
$$
is not summable as a closed form solution. However, many properties of~$D(k)$, inherited from~$N(k)$, are nevertheless clear: First, since each $N(i)$ term is a point on a bell-shaped curve, $D(k)$ must be a nondecreasing sigmoidal function. Moreover, since $N(i)$ is increasing in $\theta_0$ for all~$i$, $D(k)$ must also be increasing in $\theta_0$. This makes the importance of reducing $\theta_0$ by mitigation quite clear. That is, a reduced~$\theta_0$ leads to a smaller peak which happens earlier, and, in turn, this lowers the total number of deaths. Furthermore, it is also apparent that~$D(k)$ decreases with respect to $d_0$ for $k \le k_{N_0}$.

For $k > k_{N_0}$, the $N(i)$ terms are increasing in~$d_0$ which makes the global behaviour of $D(k)$ with respect to $d_0$ non-trivial. However, recalling that $k_{N_0}$ can be viewed as signalling the end of the pandemic, for reasonable $d_0$, the terms entering~$D(k)$ beyond $N(k_{N_0})$ will be quite small; i.e., the portion of $D(k)$ which increases with respect to $d_0$ will be small enough such that $D(k)$ decreases with $d_0$. That being said, for very large $d_0$, it is easy to see that $N(i) \approx N_0$ for all~$i$ meaning that the portion beyond $k_{N_0}$ is non-negligible, and $D(k) \approx (k+1) N_0$.


\subsection{Approximation of~$D(k)$ Using a Normal Distributions}
\vspace{-.05in}
It is also possible to enhance our understanding of total deaths via an approximation involving the classical normal distribution. To this end, we work with the continuous-time counterparts: ~$N_c(t)$ for~$N(k)$ and $D_c(t)$ for~$D(k)$. That is, beginning with the infinitesimals over time interval~$dt$
$$
\frac{dN_c}{N_c} = \frac{\log(2)}{d(t)},
$$
we integrate, substitute for~$d(t)$ and carry out a lengthy but straightforward calculation to obtain
\begin{eqnarray*}
N_c(t)& = &e^{\int_0^t \frac{\log(2)}{d(\zeta)}d\zeta}N_0\\
& = & \varphi\left(\frac{t-\mu_0}{\sigma_0}\right) C_0 N_0
\end{eqnarray*}
where
$$
C_0 \doteq 2^{\frac{\theta_0}{2d_0}}\sqrt{\frac{2\pi\theta_0d_0}{\log(2)}}
$$
is a scaling constant and~$\varphi(x)$ is the density function for a standard normal distribution; $\mu_0 \doteq \theta_0$ and $\sigma_0^2 \doteq \theta_0d_0/\log(2)$, respectively, play the roles of mean and variance for a notional ``time-to-peak'' random variable. Thus, our approximation for total deaths is given by
\begin{align*}
D_c(t) &= N_0 + \int_0^t N_c(\zeta)d\zeta \\ 
&= \left[1 + \left\{\Phi\left(\frac{t-\mu_0}{\sigma_0}\right) - \Phi\left(\frac{-\mu_0}{\sigma_0}\right)\right\}C_0\right] N_0
\end{align*}
where~$\Phi(x)$ above is the cumulative distribution function for the standard normal distribution.

\begin{figure}[htbp]
	\centering
	\includegraphics[width = 3.5in]{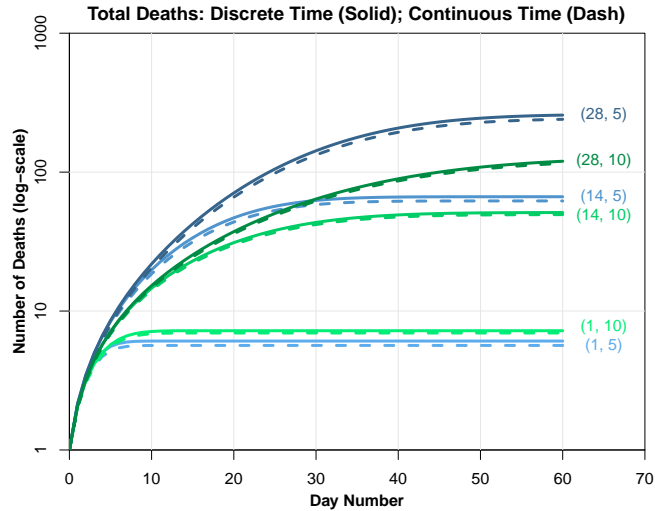}
	\caption{Total Deaths Displayed for Six $(\theta_0,d_0)$ Cases with $N_0=1$.}
	\label{fig:Dplot}	
\vspace{-.05in}
\end{figure}

To provide an indication of the quality of the above approximation, in Figure \ref{fig:Dplot}, we display $D(k)$ along with $D_c(k)$ for six $(\theta_0,d_0)$ combinations. It is clear that they are relatively close for the cases considered and we have found this to be true for a wide range of practical parameter values. Specifically, the maximum relative difference is approximately 11\% for $\theta_0,d_0>2$ and 5\% for $\theta_0,d_0>7$, but can be large if $\theta_0$ and $d_0$ are both very small, which is not likely in practice.

\subsection{Asymptotic Behavior of~$D(k)$}
\vspace{-.05in}
We now gain insight into the asymptotic total number of deaths by letting $t$ tend to infinity and obtain
$$
D_c(\infty) = \lim_{t \rightarrow \infty}D_c(t) = \left[1 + \left\{1 - \Phi\left(\frac{-\mu_0}{\sigma_0}\right)\right\}C_0\right] N_0.
$$
Of particular interest is the dependence on $d_0$ which could not be fully characterized for $D(\infty)$ previously. Thus, differentiating, we find that
$$
\frac{\partial\log D_c(\infty)}{\partial d_0} = \frac{N_0- D_c(\infty)}{2d_0 D_c(\infty) } \left\{z_0^2 + A(z_0) z_0 - 1\right\}
$$

where~$z_0 = \mu_0/\sigma_0$ and $A(z_0) = \varphi(-z_0)/(1 - \Phi(-z_0))$.
Clearly, for $\theta_0, d_0 >0$, the above derivative is negative if $z_0^2 + A(z_0) z_0 - 1 > 0$. A straightforward numerical calculation shows that this holds true for $z_0 > 0.84$ which is equivalent to $d_0 < 0.98 \theta_0$. 
Thus, $D_c(\infty)$ decreases with $d_0$ until it reaches $\theta_0$, and, although this is an approximation to what happens for $D(\infty)$, it is nonetheless a useful insight into its behavior. In practice, we have found that $\theta_0$ is larger than $d_0$; see the analysis of the two countries in Section \ref{sec:numerical} and note that this also true for a variety of other countries not shown. In summary, for practical purposes, one can view the total deaths as decreasing in $d_0$. 

\section{Model Estimation and Numerical Examples\label{sec:numerical}}
\vspace{-.05in}
Given a data set of actual deaths, $N_a(1), \ldots, N_a(n)$, for the purpose of estimating the parameters $N_0$, $d_0$, and  $\theta_0$, using the resulting total death values $D_a(k)$, we minimize
$$
\sum_{k=1}^n (D_a(k) - D(k))^2.
$$
Since this objective function in nonlinear and non-convex in the parameters, we obtain good initial values by working with log-new-deaths. Then, a straightforward calculation leads to a classical linear least squares problem; i.e., minimizing
$$
\sum_{k=1}^n\left(\frac{\log N_a(k)}{\log 2} - \beta_0 - \beta_1 k - \beta_2 k^2\right)^2
$$
yields $\hat \beta_0$, $\hat \beta_1$, and $\hat \beta_2$
from which we obtain estimates
$$
\hat N_0 = 2^{\hat\beta_0}; \;\; \hat d_0 = 1/(\hat\beta_1+\hat\beta_2);\;\; \hat\theta_0 = -(\hat\beta_1+\hat\beta_2)/(2\hat\beta_2).
$$
In some cases, these initial least squares estimates may be satisfactory solutions to the underlying problem in~$D_a(k)$ and~$D(k)$. In other cases, further iterations are needed because they may be sensitive to noisy daily death data.

\subsection{Numerical Examples: Brazil and Mexico}
\vspace{-.05in}
We now illustrate the application of our new model using historical data for year 2020 available in \texttt{https://ourworldindata.org/coronavirus}.~To this end, consider death data for Brazil and Mexico beginning on March 28, 2020, a day on which Brazil had ninety-two total deaths and Mexico had only twelve, and ending on June~21,~2020. 
For both countries we estimated the parameter triple~$(\hat N_0, \hat d_0, \hat \theta_0)$ using the procedure described above. Then we compared model-based predictions with actual death numbers and made projections on the asymptotic value~$D(\infty)$ obtained as~$k \rightarrow \infty$. The reader is reminded that an implicit assumption in our calculations is that the degree of mitigation~$u(k)$ is held constant; if a country prematurely relaxes measures such as social distancing, the model parameters should be recalibrated incorporating new data which comes to light.

\begin{figure}[htbp]
	\centering
	\includegraphics[width = 3.5in]{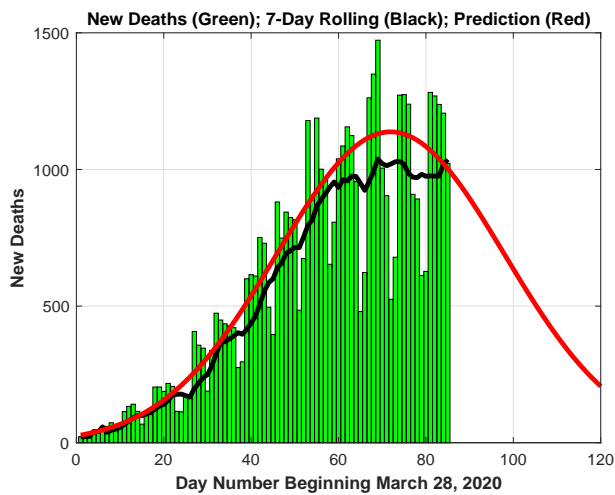}
	\caption{Daily Covid-19 Deaths in Brazil}
	\label{fig:brazil_daily}	
\vspace{-.05in}
\end{figure}

\begin{figure}[htbp]
	\centering
	\includegraphics[width = 3.5in]{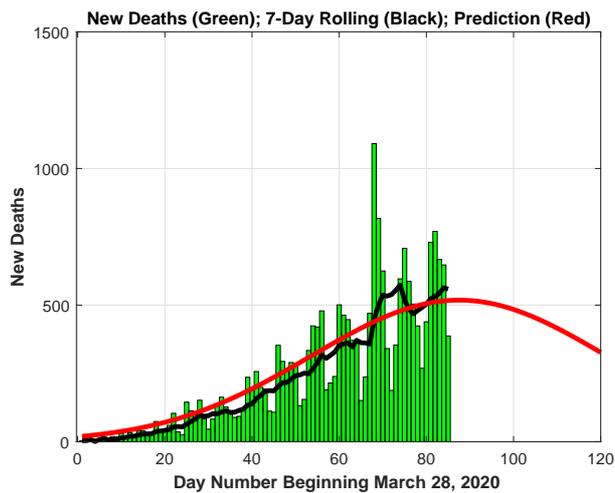}
	\caption{Daily Covid-19 Deaths in Mexico}
	\label{fig:mexico_daily}	
\vspace{-.05in}
\end{figure}

In the case of Brazil, the estimated parameter values are~$\hat N_0 = 28.0,$ $\hat d_0 = 6.68$, and $\hat \theta_0 = 70.4$. The favorable performance of our model is depicted in Figure~\ref{fig:brazil_daily} where new deaths appear to be peaking around the time the model predicts. By summing up the daily totals, we readily obtain total deaths which increase sigmoidally to~$49,976$ by the end of the observation period; interestingly, the asymptotic value~$D_\infty \approx 74,000$ is about~$50\%$ higher than this.
In the case of Mexico, the estimated parameter values are~$\hat N_0 = 19.9,$ $\hat d_0 = 9.25$, and~$\hat \theta_0 = 86.0$. The fit to the data is shown in Figure~\ref{fig:mexico_daily} where,  in contrast to Brazil, daily deaths had not quite peaked by the end of the observation period. Again by summing up new deaths, we obtain $20,781$ as the number of total deaths and note that the asymptotic value~$D(\infty) \approx 44,000$ is over~$100\%$ higher.

\section{Conclusion\label{sec:conclusions}}
\vspace{-.05in}
This paper, part of the voluminous body of literature on epidemic modelling and control, differs from previous work using SIR-type models; i.e., we do not structure the dynamics based on many possible epidemiological considerations. Rather, we focus on a death doubling parameter~$d(k)$ and a peak day parameter~$\theta(k)$, and rely on data to dynamically update the model as required. This approach to model identification is similar to those of \cite{Ma_2014} and \cite{Chowell_2017} and, as previously mentioned, is motivated by the fact that each epidemic can present a vastly different array of challenges. For a new epidemic such as Covid-19, our view is that it may be premature to use highly structured model equations which rely on detailed epidemiological factors. As evidenced by our numerical experiments, our data-driven approach, with very few parameters to be estimated, appears to fit the data quite well; this is also true for simulations we conducted for many other countries not shown.

Based on our work to date, two important directions immediately present themselves for future work: First, for the case when the mitigation level~$u(k)$ is no longer constant, it would be interest to study the evolution of deaths in an adaptive control context; i.e., as system parameters due to the arrival of new data,
the controller~$u(k)$ is correspondingly adjusted. The control-theoretic setup in Figure~\ref{fig:simulation_01} should rightfully be viewed in this more general context.

The second area for future research involves the formulation of an appropriate performance index. To provide some flavor as to the type of performance quantification issues which arise, for the current Covid-19 crisis, it is noted that societies around the world have been grappling with the following questions: How does the {\it degree of mitigation} get reflected in~$N(k)$ and~$D(k)$? What tradeoffs is a society willing to accept between ``lifestyle restrictions'' and the level of deaths? Although existing literature includes ``optimal control'' formulations for epidemics, it is silent as to the detailed construction of the performance index. To illustrate what is meant by this, for the classical quadratic case
$$
J(u) = Q\sum_{k = 1}^N N^2(k) + R\sum_{k = 0}^{N-1}u^2(k),
$$
to say the least, it may be highly challenging to choose weights~$Q$ and~$R$ reflecting the tradeoffs which a society is willing to accept. In fact, it may prove to be the case that there are other cost functionals which are better suited for the study of pandemics.

	\bibliographystyle{unsrt}

\end{document}